\documentclass[letterpaper]{article}
\usepackage{array}
\usepackage{amsmath}
\usepackage{amsfonts}
\usepackage{authblk}

\numberwithin{equation}{section}

\newcommand{\pht}{\phantom}
\newcommand{\ud}{\mathrm{d}}

\title{Pressure as a Source of Gravity}
\author{J. Ehlers}
\affil{Max Planck Institut f\"ur Gravitations Physik Golm, Germany} 
\author{I. Ozsv\'ath}
\affil{Department of Mathematics, The University of Texas at Dallas}
\author{E.L. Sch\"ucking}
\author{Y. Shang}
\affil{Department of Physics, New York University}
\date{October 1, 2005}

\begin{document}

\maketitle

\begin{abstract}
The active mass density in Einstein's theory of gravitation
in the analog of Poisson's equation in a local inertial system
is proportional to $\rho+3p/c^2$. Here $\rho$ is
the density of energy and $p$ its pressure for a perfect fluid.
By using exact solutions of Einstein's field equations in the
static case we study whether the pressure term contributes 
towards the mass.
\end{abstract}

\section{Introduction}
Mass is in Newton's theory of gravitation the source of a gravitational
field.  In a relativistic generalization of Newton's theory we expect
energy to take over this r\^ole.  But in Einstein's theory of gravitation
it is the full energy-stress tensor that becomes the source of
the gravitational field.  In the case of a perfect fluid this tensor
takes the form
\begin{equation}
\label{eq:bulk_T}
T^\mu_{\pht{\mu}\nu}=(\rho+p)u^\mu u_\nu -p \delta^\mu_\nu.
\end{equation}
Here $p$ denotes the pressure and $\rho$ the density of matter. $u^\mu$ is the 
four-velocity of matter described by a time-like unit vector with
\begin{equation}
u^\mu u_\mu=1.
\end{equation}
The speed of light $c$ has been put equal to one.

To study the r\^ole of the pressure in contributing to the gravitational field
in Einstein's theory we consider a simple case where comparison with Newtonian 
gravity is straightforward.  If we take a gravitating fluid in geodesic flow
the spatial gradient of pressure is zero and the pressure
becomes a function of time.  Such pressure has no influence on the motion
of the fluid in Newtonian dynamics.  For a local description of the flow we 
can introduce a co-moving local inertial system.  In Newton's theory the 
local acceleration of particles is then given by the negative gradient of 
a potential $V$ that itself is subject to the Poisson equation
\begin{equation}
\triangle V=4\pi G\rho,
\end{equation}
where $G$ is Newton's gravitational constant.  The local approximation to the 
relativistic description should be good since we are dealing with small 
velocities near the orgin of the co-moving
inertial system.

A local description of the flow can be obtained in Einstein's theory by using
the formulas of geodesic deviation together with the field equations.  In 
this way \cite{HS} one obtains a correction of the Poisson equation for the 
potential $V$
\begin{equation}
\label{eq:laplace}
\triangle V+\Lambda=4\pi G(\rho+3p).
\end{equation}
The correction term on the left-hand side of the equation is Einstein's 
cosmological constant that shall not be discussed further here.  On the 
right-hand side appears now the pressure contributing as source of
the gravitational field as first pointed out by Tullio Levi-Civita 
in the static case \cite{LC}.  It is easy to estimate the degree of 
this contribution.  The pressure $p$ in an ideal gas of identical 
particles with number density $n$, momentum $\vec P$ and velocity 
$\vec v$ is given by
\begin{equation}
p=\frac{1}{3}n\overline{\vec P\cdot \vec v}
\end{equation}
where the bar indicates averaging and the ``$\cdot$''
the scalar product of the two vectors.  For highly relativistic 
particles or photons this gives with $v=c$ and particles of energy $E$ 
\begin{equation}
3p=n \overline E=\rho.
\end{equation}
This result together with \eqref{eq:laplace} indicates that 
the ``active'' mass density generating a gravitational 
field for a photon gas is twice as large as one would have 
derived from Newton's theory.  In fact, if one studies cosmological 
models in Newtonian theory \cite{Mi} one obtains for a specific 
energy of $0$, $\frac{1}{2}$ or $-\frac{1}{2}$ the Friedmann 
equation for the scale factor $R(t)$ for incoherent matter, meaning 
vanishing pressure.  To obtain the Friedmann equation for a universe 
filled with radiation, like the early universe, one needs to introduce
the $3p$-term into the Poisson equation.  This was noticed by William
McCrea \cite{Mc}.

It was Richard Tolman \cite{To} who studied universes filled with 
radiation who began to wonder about the consequences of the $3p$-term 
for gravitational theory. The following scenario is known as Tolman's 
Paradox: A static spherical box has been filled with a gravitating 
substance of a given mass. If this substance undergoes an internal 
transformation (e.g. matter and anti-matter turning into radiation) 
raising the pressure, the active mass in the box would change because 
of the $3p$-term since the energy is conserved. However, such an 
internal transformation should not affect the mass measured 
outside the box, say by an orbiting particle obeying Kepler's third law. 
In a spherically symmetric field the particle should be oblivious to 
all spherically symmetric changes inside its orbit, a consequence of 
the vacuum equations known as Birkhoff's Theorem \cite{Bi}.
     
Charles Misner and Peter Putnam were intrigued by Tolman's Paradox \cite{MP}. 
They showed that by increasing the pressure inside the box one has to 
have stresses in the walls keeping the matter inside confined and 
the field static \cite{Lm}. These stresses would make negative contributions to 
the active mass that would just compensate those arising from the $3p$ 
term inside. This plausible resolution of the paradox suffered, however, 
from the restriction that the authors neglected a possible influence 
of the gravitational field on the cancellation. They solved, essentially, 
a problem in special relativity. It would, therefore, seem desirable to 
look at the problem again and study a situation where the gravitational 
field is fully taken into account.
     
To get exact and transparent results we simplify the model as follows: 
We take a sphere of fluid with constant energy density kept together by its own 
gravitation and the surface tension of a membrane under a given inside 
pressure at the surface. This boxed fluid sphere has a 
mass determined by the Schwarzschild-Droste vacuum field outside. By 
raising the surface tension of the membrane we squeeze the sphere and 
increase the pressure inside accordingly to stay in equilibrium. 
The assumption that the fluid has constant energy density simplifies all calculations 
because the radius of the sphere does not change, nor does the density 
of matter inside.  Since neither the volume nor the surface area vary,
changes in pressure and surface tension do no work and thus do not change 
the energy.

And now we want to answer the question whether the mass of the sphere 
measured outside, consisting of the mass of the fluid and the membrane, 
is affected by the raising of the pressure inside. 

To deal with this problem we are fortunate that the gravitational field 
of a fluid with constant energy density was discovered already by Karl Schwarzschild 
\cite{Sc} and is well known. Our next concern is the theory of the membrane.

\section{The Second Fundamental Form and Surface Energy-Stress Tensor}
On a time-like hypersurface the metric components must 
be continuous across the surface.  This holds also for
their derivatives along tangential directions to the hypersurface.  
However, the derivatives of the metric components in the normal 
direction to the hypersurface in general can have discontinuities. One may hope 
that an invariant formulation of the problem might be useful because one 
can then use a system with coordinate singularities for which
the calculations become simpler.

Generally speaking we are dealing with a non-null
hypersurface in a semi-Riemannian space with metric
\begin{equation}
\ud s^2=g_{\mu\nu}(x^\lambda)\ud x^\mu\ud x^\nu; 
\quad \lambda,\mu,\nu=1,\dots,n.
\end{equation}
We assume that the hypersurface is defined by
\begin{equation}
F(x^\lambda)=0
\end{equation}
with 
\begin{equation}
g^{\mu\nu}F_{,\mu}F_{,\nu}<0.
\end{equation}
This enables us to define on the hypersurface and its
neighborhood a unit space-like normal vector $n^\nu$ 
such that
\begin{equation}
n^\nu=\frac{g^{\nu\mu}F_{,\mu}}{\sqrt{|g^{\alpha\beta}F_{,\alpha}F_{,\beta}|}},
\qquad n_\nu n^\nu=-1,
\end{equation}

We introduce then a projection tensor $h^\mu_{\pht{\mu}\nu}$ by
\begin{equation}
\label{eq:shell_metric}
h^\mu_{\pht{\mu}\nu}=\delta^\mu_\nu+n^\mu n_\nu.
\end{equation}
It has the property to annihilate any vector proportional
to $n^\nu$ and to reproduce any vector orthogonal to $n^\nu$.

We form now the covariant derivative of the unit vector
$n_\nu$. First of all 
\begin{equation}
n^\nu n_{\nu;\mu}=0
\end{equation}
since $n^\nu$ is a unit vector.  We multiply the tensor
$n_{\nu;\mu}$ with the projection operator
$h^\mu_{\pht{\mu}\lambda}$ and symmetrize in the
free indices $\lambda$ and $\mu$ and we get
\begin{equation}
K_{\nu\lambda}=\frac{1}{2}(n_{\nu;\mu}h^\mu_{\pht{\mu}\lambda}
+n_{\lambda;\mu}h^{\mu}_{\pht{\mu}\nu}).
\end{equation}
It satisfies
\begin{equation}
K_{\nu\lambda}=K_{\lambda\nu}, \quad K_{\nu\lambda}n^\lambda=0, 
\quad K_{\nu\lambda}n^\nu=0.
\end{equation}
This symmetrical tensor with tangent vector $X^\nu$ of 
the hypersurface gives the 2nd fundamental form $II$
\begin{equation}
II\equiv K_{\nu\lambda} X^\nu X^\lambda.
\end{equation}
If we have a boundary surface which carries an energy-stress tensor 
with a delta function on it, the tensor given above can have a jump 
between the two sides of this surface.  The Lanczos-Israel junction 
condition \cite{Lanc,Is}
gives the relation between the surface energy-stress tensor 
$I_{\mu\nu}$ and the jump of the 2nd fundamental form as
\begin{equation}
\label{eq:Israel_condition}
\kappa I_{\mu\nu}=K_{\mu\nu}\big|^+_-
-h_{\mu\nu} K^\lambda_\lambda \big|^+_-, \qquad \kappa=8\pi G,
\end{equation} 
where all values are calculated on the boundary.  
The signs $+$ and $-$ denote the values calculated on the 
positive and negative sides (with respect to $n^\nu$)
of the surface respectively and a vertical line denotes the subtraction
between the two values.  We will use these notations
again later.

A translation into English of the original Lanczos paper 
\cite{Lanc} can be found in  Cornelius Lanczos ``Collected 
Published Papers with Commentaries'' \cite{col2}. 
The useful commentary by J. David Brown
\cite{Br} lists the further 
references \cite{Ip,LM} to which we might add 
\cite{Ba,CO,MK}.  See also Appendix \ref{app:boundary}.

\section{The Empty Bubble}
For orientation we treat first the equilibrium of a spherical,
gravitating empty bubble with a constant surface stress $\tau$ in 
Newtonian theory.  We count $\tau$ positive if it acts
as a surface tension that tries to minimize the surface area.
As a compressive stress $\tau<0$ takes negative values.
We call the radius of the bubble $r_0$ and assume that 
matter is distributed on its surface with constant mass density $\sigma$.
The total mass $M_S$ of the bubble, that is the sum of its differential 
masses, is given by
\begin{equation}
M_S=4\pi\sigma r_0^2.
\end{equation}
A surface element of area $\ud A$ is attracted toward the center of 
the sphere by a force
\begin{equation}
\label{eq:newtonian_F}
\ud F=-G \frac{M_S \sigma\ud A}{2r_0^2}=-2\pi G \sigma^2\ud A.
\end{equation}
It is the mean of the attraction just outside and
just inside of the membrane.  The surface stress $\tau$ acts 
on the surface element $\ud A$ with an outward directed 
radial force $\ud F'$
\begin{equation}
\ud F'=-\frac{2 \tau \ud A}{r_0}.
\end{equation}
This formula is easily derived by considering $\ud A$ as a circular disc
on the sphere with the infinitesimal radius $\epsilon r_0$. The area of this 
disc is then given by $\pi (\epsilon r_0)^2$. The radial component of 
the surface stress $\tau$ is $\epsilon \tau$. Integrating it 
along the circumference of the circle amounts to multiplying it with 
$2 \pi \epsilon r_0$. We have thus for the circular surface element the 
radial force
\begin{equation}
\ud F'=-(2\pi \epsilon r_0) \epsilon \tau=-\frac{2 \tau \ud A}{r_0}.
\end{equation}
The formula is a special case of the formula derived for the theory 
of capillarity by Thomas Young \cite{Yo} in 1805 and also a year later by 
Pierre Laplace \cite{La} to whom it is usually attributed. Instead of the 
factor $2/r_0$ the general formula has the sum of the principal curvatures
that coincide for the sphere in our case.

Adding $F$ and $F'$ to zero for equilibrium gives
\begin{equation}
\label{eq:newtonian_tau}
-\tau=\pi G \sigma^2 r_0=\frac{GM_S\sigma}{4 r_0},
\end{equation}
from which we have that condition $\sigma>|\tau|$ is equivalent to
\begin{equation}
\frac{GM_S}{r_0}<4.
\end{equation}
The gravitational binding energy of the bubble is given by
\begin{equation}
E_{pot}=-\frac{G M_S^2}{2r_0}.
\end{equation}
As a relativistically corrected mass $M$ of the bubble
we might surmise
\begin{equation}
\label{eq:active_mass_newtonian}
M=M_S-\frac{GM_S^2}{2r_0}.
\end{equation}

We turn now to the relativistic treatment of such a model.  
It is known that outside of the sphere the space-time has 
a Schwarzschild geometry and inside it is a flat Minkowki space.  
Therefore, in the interior $r\le r_0$ we have the metric
\begin{equation}
\ud s^2=\ud \tilde t^2-\ud \tilde r^2-\tilde r^2(\ud\theta^2+\sin^2\theta\ud\phi^2),
\end{equation}
where we have used special notations for time coordinate $\tilde t$ and radial 
coordinate $\tilde r$ because they will be further adjusted soon
to match the metric components across the boundary surface.
In the exterior $r>r_0$ we take the Schwarzschild-Droste metric
\begin{equation}
\ud s^2=\left(1-\frac{2m}{r}\right)\ud t^2-\frac{\ud r^2}{1-2m/r}
-r^2(\ud\theta^2+\sin^2\theta \ud \phi^2).
\end{equation}
Matching the metric component $g_{00}$ at $r=r_0$ we find that
if $\tilde t=\alpha t$ 
\begin{equation}
\alpha^2=1-\frac{2m}{r_0}.
\end{equation}
Matching $g_{22}$ and $g_{33}$ at $r=r_0$ leads to the conclusion
that if $\tilde r$ is a smooth function of $r$ it must have the 
property 
\begin{equation}
\tilde r(r_0)=r_0.
\end{equation}
Define $\tilde r'=\ud \tilde r/\ud r$ and $\beta \equiv \tilde r'(r_0)$.
Matching $g_{11}$ gives 
\begin{equation}
\beta=\alpha^{-1}.
\end{equation}
Therefore, the interior metric becomes
\begin{equation}
\ud s^2=\alpha^2\ud t^2-\tilde r'^2\ud r^2
-\tilde r^2(\ud\theta^2+\sin^2\theta\ud\phi^2),
\end{equation}
where $\tilde r$ and $\tilde r'$ are understood as functions of $r$.

We want to calculate the tensor $K_{\mu\nu}$ 
of the hypersurface $r=r_0$ on both sides.
To do so we define a unit normal vector
in the interior as
\begin{equation}
n^\nu=\delta^\nu_1\tilde r'^{-1}
\end{equation}
and in the exterior as
\begin{equation}
n^\mu=\delta^\mu_1\sqrt{1-2m/r}.
\end{equation}
On the boundary they are both
\begin{equation}
n^\mu(r_0)=\delta^\mu_1\sqrt{1-2m/r_0}.
\end{equation}

The covariant derivative of $n_\nu$ is given by
\begin{equation}
n_{\nu;\mu}=n_{\nu,\mu}-n_\alpha \Gamma^\alpha_{\nu\mu}.
\end{equation}
The last term can be written as
\begin{equation}
-\Gamma^\alpha_{\nu\mu}n_\alpha
=-\Gamma_{\nu\mu,\alpha}n^\alpha=-\Gamma_{\nu\mu,1}n^1
=\frac{n^1}{2}(g_{\nu\mu,1}-g_{1\nu,\mu}-g_{1\mu,\nu}).
\end{equation}
On both sides of the surface, the tensor
$K_{\mu\nu}$ can be expressed as
\begin{equation}
\label{eq:2nd_fundamental_form}
K_{\mu\nu}=\frac{n^1}{2}
\left(\begin{array}{cccc}
g_{00,1}&0&0&0\\
0&0&0&0\\
0&0&g_{22,1}&0\\
0&0&0&g_{33,1}
\end{array}\right).
\end{equation}
We will use this expression to calculate our second 
fundamental forms.  For the current example, 
straightforward calculations show that
\begin{equation}
g_{00,1}\big|^+_-=\frac{2m}{r_0^2}
\end{equation}
and
\begin{equation}
g_{22,1}\big|^+_-=-2r_0(1-\beta), \qquad 
g_{33,1}\big|^+_-= g_{22,1}\big|^+_- \sin^2\theta.
\end{equation}
Therefore, we have
\begin{equation}
\begin{split}
K_{00}\big|^+_-=&\frac{m}{r_0^2}\sqrt{1-2m/r_0}\\
K_{22}\big|^+_-=&r_0\left(1-\sqrt{1-2m/r_0}\right)\\
K_{33}\big|^+_-=&K_{22}\big|^+_-\sin^2\theta.
\end{split}
\end{equation}

Using the Lanczos-Israel condition \eqref{eq:Israel_condition} and
writing
\begin{equation}
I^\mu_{\pht{\mu}\nu}=\left(
\begin{array}{cccc}
\sigma &0&0&0\\
0&0&0&0\\
0&0&\tau&0\\
0&0&0&\tau
\end{array}\right),
\end{equation}
we find
\begin{equation}
\label{eq:surface_density_empty}
\kappa\sigma=\frac{2}{r_0} \left(1-\sqrt{1-2m/r_0}\,\right)
\end{equation}
and
\begin{equation}
\label{eq:surface_stress_empty}
\begin{split}
-\kappa\tau=&\frac{1}{r_0}\left(\sqrt{1-2m/r_0}-1
+\frac{m/r_0}{\sqrt{1-2m/r_0}}\right)\\
=&\frac{1}{r_0}\left(\frac{1-m/r_0}{\sqrt{1-2m/r_0}}-1\right)
=\frac{1}{2r_0}\frac{\left(\sqrt{1-2m/r_0}-1\right)^2}
{\sqrt{1-2m/r_0}},
\end{split}
\end{equation}
or with \eqref{eq:surface_density_empty}
\begin{equation}
-\tau=\frac{\kappa \sigma^2 r_0}{8-4\kappa\sigma r_0}
=\frac{\pi G r_0\sigma^2}{1-4\pi G r_0\sigma}.
\end{equation}
Here $\tau<0$ requires that $4\pi G r_0\sigma<1$.  Solving equation 
\eqref{eq:surface_density_empty} for $m$ leads to 
\begin{equation}
m=\frac{\kappa\sigma r_0^2}{2}-\frac{(\kappa\sigma)^2r_0^3}{8}.
\end{equation}
If we define the surface internal energy as
\begin{equation}
M_S\equiv 4\pi r_0^2\sigma
\end{equation}
the active mass measured from outside $M=m/G$ is given by
\begin{equation}
M=M_S-\frac{GM_S^2}{2r_0},
\end{equation}
which is what we suggested in equation 
\eqref{eq:active_mass_newtonian}.  Here $GM_S/2r_0<1$.

The mass $M$ is given by Edmund Whittaker \cite{EW, Ko, Eh, Wa} as
\begin{equation}
M=\int\sqrt{g_{00}}(2\rho-T)\ud^3 V. 
\end{equation}
Applied to our bubble this gives
\begin{equation}
M=4\pi r_0^2\sqrt{g_{00}(r_0)}(\sigma-2\tau)
=8\pi\frac{m}{\kappa}=\frac{m}{G},
\end{equation} 
and it agrees with what we just found.
Finally we point out that with these notations
\begin{equation}
\label{eq:surface_tension_empty}
-\tau=\frac{GM_S\sigma/r_0}{4(1-GM_S/r_0)}.
\end{equation}
The condition of ``energy-dominance'' $|\tau|<\sigma$ requires that
\begin{equation}
\frac{GM_S}{r_0}<\frac{4}{5}.
\end{equation}
In the limit $m/r_0\ll1$ we find
\begin{equation}
-\tau=\frac{GM_S\sigma}{4r_0}.
\end{equation}
This gives the correct Newtonian limit of the surface tension as
we mentioned above.  These results were obtained by Kornel
Lanczos \cite{Lanc}.

\section{The Squeezed Ball}
We wish to study the following model in Einstein's theory of gravitation. 
We take a spherical ball of constant energy density. In its interior the 
ball is kept in equilibrium by its pressure gradient balancing the 
gravitational pull of the underlying matter. To simplify the following 
we assume that the energy density is independent of the pressure.  Although 
incompressibility is not an allowed material property since it would lead to
an infinite adiabatic speed of sound, constancy of energy density is 
nevertheless possible in special configurations as discussed by 
Christian M{\o}ller \cite{Mo}.  At the surface of the ball where the 
pressure is positive or zero we have a membrane. For vanishing 
surface pressure such a membrane is not necessary, but for a positive 
pressure at the surface the membrane is there for keeping the ball 
in equilibrium. 

The membrane acts in two ways. Its mass exerts a pressure on the ball
and tension squeezes the ball.  Since $\rho=\textrm{const.}$ we also 
have to assume that 
$\sigma=\textrm{const.}$.

To understand the gravitating r\^ole of pressure itself --- not of pressure 
gradients --- we want to squeeze the ball while keeping it in equilibrium. In 
this way we raise the overall pressure in the ball and compensate its rise 
at the surface by increasing the tension in the membrane. To effect such a 
process one could imagine to lower masses symmetrically and infinitely slowly 
from all sides to rest on the surface to increase the pressure. The additional 
weights are then lifted again and replaced by the increased surface tension in 
the membrane. No work on the ball will be done if its mass before and after 
is the same.

It is clear that in carrying out these internal transformations in a 
spherically symmetric fashion we are severely constrained by Birkhoff's
theorem.  We should imagine that all our machinery needed for
lowering and raising weights has to be inside a spherical
shell used as a scaffold for these operations that cannot change
the total mass (that of the scaffold included) measured at 
$r\rightarrow\infty$.

The ball itself is described by the metric of the interior Schwarzschild
solution \cite{Sc}, given for
$r\le r_0$ by
\begin{equation}
\label{eq:metric_interior_smooth}
\ud s^2=\frac{1}{4}(3a_0-a)^2\ud t^2-\frac{\ud r^2}{a^2}
-r^2(\ud\theta^2+\sin^2\theta\ud \varphi^2)
\end{equation}
with 
\begin{equation}
\label{eq:a_smooth}
a=\sqrt{1-r^2/R^2}\quad \textrm{and} \quad a_0=
\sqrt{1-r_0^2/R^2}.
\end{equation}
The energy density $\rho$ is given by
\begin{equation}
\label{eq:energy_density}
\kappa\rho=\frac{3}{R^2}=\textrm{const.},\qquad
\kappa=8\pi G.
\end{equation}
The constant $R$ is the radius of curvature of
$3$-space of constant positive curvature described by
$t=\textrm{const}$.  The pressure $p$ inside the
ball is
\begin{equation}
\label{eq:pressure_smooth}
\kappa p=\frac{3(a-a_0)}{R^2(3a_0-a)}=\kappa\rho\frac{a-a_0}{3a_0-a}.
\end{equation}
At the surface of the ball at $r=r_0$ the pressure vanishes.  The central
pressure $p(0)$ is given by
\begin{equation}
\kappa p(0)=\frac{3(1-a_0)}{R^2(3a_0-1)}.
\end{equation}
This introduces the limitation
\begin{equation}
\label{eq:pressure_condition}
a_0>\frac{1}{3},\quad \textrm{or} \quad \frac{r_0}{R}<\frac{\sqrt 8}{3}.
\end{equation}

For $r\ge r_0$ we use the Schwarzschild-Droste vacuum solution
\begin{equation}
\label{eq:metric_exterior_smooth}
\ud s^2=\left(1-\frac{2m}{r}\right)\ud t^2-\frac{\ud r^2}{1-2m/r}
-\ud r^2(\ud\theta^2+\sin^2\theta\ud\varphi^2).
\end{equation}
We fit the two metrics \eqref{eq:metric_interior_smooth}
and \eqref{eq:metric_exterior_smooth} continuously together
at $r=r_0$ by putting
\begin{equation}
1-\frac{2m}{r_0}=a(r_0)^2=1-\frac{r_0^2}{R^2}.
\end{equation}
With \eqref{eq:energy_density}, this equation relates the mass 
$M=m/G$ to the energy density as
\begin{equation}
\label{eq:schwarzschild_mass}
m=GM=\frac{r_0^3}{2R^2}=\frac{\kappa r_0^3\rho}{6}=G\frac{4\pi}{3}r_0^3\rho.
\end{equation}
Therefore, if we define
\begin{equation}
\label{eq:def_volume_mass}
M_V\equiv\frac{4\pi r_0^3}{3}\rho,
\end{equation}
we have
\begin{equation}
M=M_V.
\end{equation}
The Euclidean volume appearing on the right-hand
side of equation \eqref{eq:def_volume_mass} is not the true volume 
in the space of constant positive curvature.  The volume element of
a spherical shell of thickness $\ud r$ is given by
\begin{equation}
\label{eq:volume}
\ud V=\frac{4\pi r^2}{a}\ud r.
\end{equation}
As shown by Edmund Whittaker \cite{EW} the mass generating the 
gravitational field, is obtained for static fields by the integral
\begin{equation}
\label{eq:mass_integral_smooth}
M=\int_V\sqrt{g_{00}}(\rho+3p)\ud V
\end{equation}
and it agrees with $M_V$.  Besides the addition of
$3p$ to the density and the modified volume element,
the factor $\sqrt{g_{00}}$ appears that represents
the gravitational potential in the weak field approximation.  Since
we have for the interior Schwarzschild solution that
\begin{equation}
\sqrt{g_{00}}(\rho+3p)=\rho a
\end{equation}
the integral \eqref{eq:mass_integral_smooth} will give
the Euclidean value with \eqref{eq:volume}.

Since we assumed that the pressure vanished at $r=r_0$
we should also have continuity of the first derivatives
of the two metrics \eqref{eq:metric_interior_smooth}
and \eqref{eq:metric_exterior_smooth}.  This is not the
case in the coordinates used.  However, if
the matching problem is formulated in terms of
the second fundamental forms at the
hypersurface $r=r_0$ we shall see later that the matching
conditions are fulfilled.

We now want to put a delta-like membrane on the boundary
$r=r_0$.  The membrane, in general, has a non-vanishing 
energy-stress tensor on it, and therefore the interior of 
the bubble might be squeezed and its pressure increases.  
If the energy density of the fluid remains unchanged, we can 
take the interior metric to be
\begin{equation}
\ud s^2=\frac{1}{4}(3a_1-a)^2\ud \tilde t^2-\frac{\ud \tilde r^2}{a^2}
-\tilde r^2(\ud\theta^2+\sin^2\theta\ud \varphi^2)
\end{equation}
with 
\begin{equation}
a=\sqrt{1-\tilde r^2/R^2}\quad \textrm{and} 
\quad a_1=\sqrt{1-r_1^2/R^2},
\end{equation}
where $r_1>r_0$.  Again we have defined new coordinates
$\tilde t$ and $\tilde r$ just so that we can adjust them to 
match the outside metric.  The constant density remains as
\begin{equation}
\kappa\rho=\frac{3}{R^2}.
\end{equation}
and the pressure becomes
\begin{equation}
\label{eq:pressure}
\kappa p=\frac{3(a-a_1)}{R^2(3a_1-a)}=\kappa\rho\frac{a-a_1}{3a_1-a},
\end{equation}
which no longer vanishes at the boundary $r=r_0$.
It follows from \eqref{eq:pressure} that
\begin{equation}
r_1 <\frac{\sqrt 8}{3}R
\end{equation}
has to hold to avoid that the pressure becomes infinite at the center
$\tilde r=0$.  In the exterior we assume a Schwarzschild-Droste metric
\begin{equation}
\ud s^2=\left(1-\frac{2m'}{r}\right)\ud t^2
-\frac{\ud r^2}{1-2m'/r}
-r^2(\ud\theta^2+\sin^2\theta \ud\varphi^2)
\end{equation}
with a parameter $m'$ to be determined later.

Again matching $g_{22}$ and $g_{33}$ dictates that
$\tilde r$ as a function of $r$ must have the property that
\begin{equation}
\tilde r(r_0)=r_0.
\end{equation}
The continuity of $g_{11}$ determines that
\begin{equation}
\label{eq:mass_condition}
1-\frac{2m'}{r_0}=\frac{a_0^2}{\beta^2}
\end{equation}
which can also be expressed as
\begin{equation}
m'=\frac{r_0}{2}\left(1-\frac{a_0^2}{\beta^2}\right).
\end{equation}
Here $\beta\equiv\frac{\ud\tilde r}{\ud r}|_{r=r_0}$ is 
defined in the same way as before and $a_0$ is given
by \eqref{eq:a_smooth}.

Continuity in $g_{00}$ is achieved by matching
the time coordinates $t$ and $\tilde t$ at $r=r_0$
by putting $\tilde t=\alpha t$.  This determines $\alpha$ to be
\begin{equation}
\alpha=\frac{2\sqrt{1-2m'/r_0}}
{3a_1-a_0}=\frac{2 a_0}
{\beta(3 a_1-a_0)}.
\end{equation}
We should from now on take the interior metric as
\begin{equation}
\label{eq:metric_interior}
\ud s^2=\frac{\alpha^2}{4}
(3 a_1 - a)^2 \ud t^2
-\frac{\tilde r'^2 \ud r^2}{a^2}
-\tilde r^2(\ud\theta^2+\sin^2\theta\ud \varphi^2).
\end{equation}
where $\tilde r$ is understood as a function of $r$.  In this way
both the coordinates and the metric are continuous across 
the boundary surface.

If we call $p_s\equiv p(r_0)$ the pressure at the surface $r=r_0$,
\begin{equation}
\label{eq:surface_pressure}
\kappa p_s=\frac{3(a_0-a_1)}
{R^2(3 a_1-a_0)}
=\frac{\beta\alpha-1}{R^2},
\end{equation}
or
\begin{equation}
\label{eq:surface_pressure_alpha}
p_s=\frac{\beta\alpha-1}{3}\rho.
\end{equation}

To relate the coefficents $\alpha$ and $\beta$ to the surface energy 
density and pressure we must again calculate the second fundamental 
form of the hypersurface $r=r_0$ on either side.  We define a
unit normal vector in the interior 
\begin{equation}
n^\mu=\delta^\mu_1\tilde r'^{-1} a
\end{equation}
and in the exterior  
\begin{equation}
n^\mu=\delta^\mu_1\sqrt{1-2m'/r}.
\end{equation}
The vectors coincide on the boundary $r=r_0$.
Clearly, equation \eqref{eq:2nd_fundamental_form}
is still valid.  We have in the exterior
\begin{equation}
g^+_{00,1}\big|_{r=r_0}=\frac{2m'}{r_0^2}=\frac{1}{r_0}
-\frac{a_0^2}{r_0\beta^2} 
\end{equation}
and in the interior
\begin{equation}
g^-_{00,1}\big|_{r=r_0}=\frac{\alpha^2}{4}
\left. \frac{
\ud (3 a_1-a)^2}
{\ud \tilde r}\cdot \frac{\ud \tilde r}{\ud r}
\right|_{r=r_0}=\frac{\alpha r_0}{R^2}.
\end{equation}
Therefore 
\begin{equation}
K_{00}\big|^+_-=\frac{\sqrt{1-2m'/r_0}}{2}
\left(\frac{1}{r_0}-\frac{\alpha r_0}{R^2}
-\frac{a_0^2}{r_0\beta^2}
\right).
\end{equation}
Jumps of $K_{22,1}$ and $K_{33,1}$ are easily found
to be
\begin{equation}
K_{22}\big|^+_-=-r_0\sqrt{1-2m'/r_0}(1-\beta),\qquad 
K_{33}\big|^+_-=K_{22}\big|^+_-\sin^2\theta.
\end{equation}

Using again Lanczos-Israel junction condition 
\eqref{eq:Israel_condition} and writing 
\begin{equation}
I^\mu_{\pht{\mu}\nu}=\left(\begin{array}{cccc}
\sigma&0&0&0\\
0&0&0&0\\
0&0&\tau&0\\
0&0&0&\tau
\end{array}\right)
\end{equation}
we find with \eqref{eq:mass_condition}
\begin{equation}
\label{eq:surface_density}
\kappa \sigma=\frac{2 a_0}{r_0}
\left(1-\frac{1}{\beta}\right)
=\frac{2}{r_0}\left[a_0-\sqrt{1-2m'/r_0}\right],
\end{equation}
which implies that
\begin{equation}
\label{eq:inverse_beta}
\frac{1}{\beta}=1-\frac{\kappa\sigma r_0}{2 a_0},
\end{equation}
and 
\begin{equation}
\label{eq:surface_stress}
\begin{split}
-\kappa \tau =&\frac{1}{2\sqrt{1-2m'/r_0}}\left(
\frac{1}{r_0}-\frac{\alpha r_0}{R^2}-\frac{a_0^2}{r_0\beta^2}\right)
-\frac{\kappa\sigma}{2}.
\end{split}
\end{equation}
 
We are now ready to calculate $m'$ following a procedure
similar to the one employed in the calculation for the empty bubble.
Solving equation \eqref{eq:surface_density} for $m'$ we find that
\begin{equation}
m'=\frac{r_0^3}{2R^2}+\frac{r_0^2\kappa a_0 \sigma}{2}-
\frac{r_0^3(\kappa \sigma)^2}{8}.
\end{equation}
Here $m'$ is determined by the mass distributions,
namely $\rho$, $r_0$ and $\sigma$ and does not depend in addition
on pressure $p$ and surface tension $\tau$.  This conclusion
contains in essence already our result:`` Internal changes
of the pressure and surface stress distribution at fixed $\rho$,
$r_0$ and $\sigma$ do not change the total energy.
This means that the active mass $M$ measured from outside is
\begin{equation}
\label{eq:active_mass}
\begin{split}
M=&\frac{m'}{G}=\frac{4\pi r_0^3}{3}\rho
+4\pi r_0^2 a_0 \sigma-\pi \kappa r_0^3 \sigma^2\\
=&M_V+a_0 M_S-\frac{G M_S^2}{2 r_0}.
\end{split}
\end{equation}
The first term is clearly the contribution from the bulk energy density
and the second term is the surface mass scaled by a factor
due to the space-time curvature. The fact that it is smaller than
$M_S$ can be understood as the effect of the binding energy between 
the surface mass and the bulk mass.  It's more easily seen in the
limit $r_0/R\ll1$ where to the first order of $r_0/R$ we have
\begin{equation}
M\approx M_V+M_S-\frac{G M_S M_V}{r_0}-\frac{G M_S^2}{2r_0}.
\end{equation}
A term of Newtonian potential energy between the surface
mass and bulk mass shows up explicitly.  The last term of the 
active mass that is quadratic in $M_S$ comes
from the self gravitational binding energy of the surface as
we explained before.  In the limit
$R\rightarrow +\infty$ and therefore 
$\rho\rightarrow 0$ the active massive simplifies to 
\begin{equation}
M=M_S-\frac{GM_S^2}{2r_0}
\end{equation}
which is exactly what we've found already in the previous
section.  

These results also enable us to write down the surface stress 
in term of physical quantities.  Using equation \eqref{eq:surface_density}
we find
\begin{equation}
\sqrt{1-2m'/r_0}=a_0-\frac{\kappa\sigma r_0}{2}.
\end{equation}
This shows that 
\begin{equation}
\kappa\sigma<\frac{2 a_0}{r_0}
\end{equation}
is necessary to prevent collapsing.  Equation \eqref{eq:surface_pressure_alpha} 
and \eqref{eq:inverse_beta} lead to
\begin{equation}
\alpha=\left(\frac{3p_s}{\rho}+1\right)
\left(1-\frac{\kappa\sigma r_0}{2 a_0}\right).
\end{equation}
Inserting them into \eqref{eq:surface_stress} eventually gives
\begin{equation}
\label{eq:surface_tension_full}
\begin{split}
-\kappa\tau=&\frac{1}{2 a_0-\kappa\sigma r_0}
\left[\frac{1}{r_0}-\frac{\alpha r_0}{R^2}-\frac{a_0^2}{r_0} 
\left(1-\frac{\kappa\sigma r_0}{a_0}
+\frac{(\kappa\sigma)^2 r_0^2}
{4a_0^2}\right)\right]-\frac{\kappa\sigma}{2}\\
=&\frac{1}{2 a_0-\kappa\sigma r_0}
\left[\frac{r_0}{R^2}(1-\alpha)
+\kappa a_0 \sigma 
-\frac{(\kappa\sigma)^2 r_0}{4}\right]-\frac{\kappa\sigma}{2}\\
=&-\frac{\kappa r_0 p_s}{2 a_0}+
\frac{\kappa}{2 a_0-\kappa\sigma r_0}
\left[\frac{\sigma}{a_0}\left(1-\frac{r_0^2}{2R^2}\right)
-\frac{\kappa\sigma^2 r_0}{4}\right]-\frac{\kappa\sigma}{2}\\
=&-\frac{\kappa r_0 p_s}{2 a_0}+
\frac{\kappa}{2 a_0-\kappa\sigma r_0}
\left[\frac{\sigma r_0^2}{2a_0 R^2}
+\frac{\kappa\sigma^2 r_0}{4}\right]\\
=&-\frac{\kappa r_0 p_s}{2 a_0}+
\frac{\kappa G \sigma}{r_0(a_0-\kappa\sigma r_0/2)}
\left(\frac{M_V}{2a_0}
+\frac{M_S}{4}\right).
\end{split}
\end{equation}

As a check on this equation we put the surface pressure
$p_s=0$, and remove the mass from the interior
of the bubble, i.e., putting also $M_V=0$.
In this case we obtain for the surface stress $\tau$
of the empty bubble our previous result from equation
\eqref{eq:surface_tension_empty}
\begin{equation}
\tau=\frac{GM_S\sigma/r_0}{4(1-GM_S/r_0)}.
\end{equation}
For the unsqueezed ball with vanishing surface pressure
and vanishing surface energy density, $p_s=0$ and $\sigma=0$,
we find, as expected, that the surface stress $\tau$ vanishes.

We have reached here the main objective of this paper.  We wish
to demonstrate explicitly the cancellation between the contributions 
of pressure $p(r)$ and surface tension $\tau$ as a source of the 
gravitational field.  This can only be achieved by calculating $\tau$ 
in terms of other physical quantities first as we have done above.  
We can now calculate the active mass $M$ by the following integration.  
First we notice that 
\begin{equation}
\sqrt{g_{00}}\,(\rho+3p)=\rho \alpha a.
\end{equation}
Therefore, the active mass can be evaluated as
\begin{equation}
\begin{split}
M=&\int_V\sqrt{g_{00}} (\rho+3p) \ud V
+\sqrt{g_{00}(r_0)}r_0^2\int_S (2\tau+\sigma)\ud\Omega\\
=&\frac{4\pi r_0^3}{3}\alpha\rho
+4\pi r_0^2\left[\frac{r_0}{3}\rho(1-\alpha)
+\sigma a_0
-\frac{\kappa\sigma^2 r_0}{4}\right]\\
=&\frac{4\pi r_0^3}{3}\rho+4\pi r_0^2\sigma a_0
-\pi\kappa r_0^3\sigma^2 \\
=&M_V+a_0 M_S-\frac{GM_V M_S}{r_0},
\end{split}
\end{equation}
where $\ud\Omega=\sin\theta\ud\theta\ud\varphi$.
This coincides with \eqref{eq:active_mass} as it must and we can see 
the intriguing cancellations in the second step above.

\section{The Case of a Massless Membrane}
We first wish to consider the case of a ball of constant 
density $\rho$ and radius $r_0$ under its own gravitation. Density
and radius of this spherical ball can be chosen
freely.  The pressure $p_s=p(r_0)$ at the surface vanishes.
and $p(r)$ is given by \eqref{eq:pressure_smooth}
and \eqref{eq:a_smooth}
\begin{equation}
p(r)=\rho \frac{\sqrt{1-r^2/R^2}-\sqrt{1-r_0^2/R^2}}
{3\sqrt{1-r_0^2/R^2}-\sqrt{1-r^2/R^2}}.
\end{equation}
This gives for small values of $r_0 \ll R$
\begin{equation}
\label{eq:dis_pressure}
p(r)=\rho(r_0^2-r^2)/4R^2.
\end{equation}
The Schwarzschild mass $M$ from \eqref{eq:schwarzschild_mass}
is given by
\begin{equation}
\label{eq:dis_mass}
M=\frac{4\pi}{3}\rho r_0^3.
\end{equation}

If we squeeze this ball by surrounding it by a massless 
($\sigma=0$) membrane with surface tension $\tau$ it will develop
a pressure inside given by \eqref{eq:pressure} 
\begin{equation}
\label{eq:dis_squeezed_pressure}
p_1(r)=\rho \frac{\sqrt{1-r^2/R^2}-\sqrt{1-r_1^2/R^2}}
{3\sqrt{1-r_1^2/R^2}-\sqrt{1-r^2/R^2}}, \qquad r_1 \ge r_0.
\end{equation}
$M$ remains the same according to \eqref{eq:active_mass} since $M_S=0$.
On the surface this pressure is given by
\begin{equation}
p_s=p_1(r_0)=\rho \frac{\sqrt{1-r_0^2/R^2}-\sqrt{1-r_1^2/R^2}}
{3\sqrt{1-r_1^2/R^2}-\sqrt{1-r_0^2/R^2}}.
\end{equation}
The pressure at the surface is matched by the surface tension
$\tau$ according to \eqref{eq:surface_tension_full} 
\begin{equation}
\tau=r_0 p_s/2\sqrt{1-2GM/r_0}.
\end{equation}
For small values of $r_0\ll R$ the pressure $p_1(r)$ is according to
\eqref{eq:dis_squeezed_pressure}
\begin{equation}
p_1(r)=\rho(r_1^2-r^2)/4R^2.
\end{equation}
On the surface for $r=r_0$ this is
\begin{equation}
\label{eq:dis_surface_pressure}
p_s=\rho(r_1^2-r_0^2)/4R^2.
\end{equation}
Comparing \eqref{eq:dis_pressure} with \eqref{eq:dis_surface_pressure}
we see that
\begin{equation}
p_1(r)-p(r)=p_s
\end{equation}
for small values of $r_0/R$.  Under these conditions the raising
of a surface tension in the membrane results in a constant increase
of pressure by $p_s$ inside the whole sphere.  If $r_0/R$ becomes
comparable to $1$ the pressure towards the center increases
faster than $p_s$ as is easy to derive from \eqref{eq:dis_squeezed_pressure}.

We can now consider $\tau$ or  $p_s$ as a new independent parameter
for our model.  The crucial result of our investigation is
that the Schwarzschild mass $M$ of \eqref{eq:dis_mass}
is independent of the surface pressure $p_s$.  While
the density of the active mass inside the sphere depends on the surface
pressure $p_s$ it does not influence the mass measured from outside for
a system in a static equilibrium.  

The case where we remove the unphysical assumption that the membrane
be massless ($\sigma=0$) really shows nothing essentially new.  
Equation \eqref{eq:surface_tension_full} relating surface tension
$\tau$ with surface pressure $p_s$ becomes more complicated as
does the active mass in \eqref{eq:active_mass}.  But the
conclusion remains the same that $M$ is independent of the 
surface pressure $p_s$ in equilibrium.

It would seem, therefore, that the $3p$ addition to the
active mass density that arose in Tolman's paradox would be
compensated by the forces that are responsible for
equilibrium.  We do not expect that calculations with a more
realistic equation of state (we used incompressibility) and a different
scenario in raising pressure would change the conclusions.  If one wants
to find the effect of the $3p$-term one has to look at
non-equilibrium situations as they present themselves in the early
universe or in the last stages of a type II supernova core.

\appendix

\section*{Appendix}

\section{}
\label{app:boundary}
The total energy-momentum tensor $T^\mu_{\pht{\mu}\nu}$ for a bulk
fluid and a membrane can be wirtten as
\begin{equation}
T^\mu_{\pht{\mu}\nu}=\rho u^\mu u_\nu-p(\delta^\mu_\nu-u^\mu u_\nu)
+\{\sigma u^\mu u_\nu-\tilde p (h^\mu_{\pht{\mu}\nu}- u^\mu u_\nu)\}\delta(n)
\end{equation}
The bulk part of the tensor was given in equation \eqref{eq:bulk_T}.
The shell part $I^\mu_{\pht{\mu}\nu}$ is constructed in analogy to
the bulk part with surface mass density $\sigma$ and surface pressure
$\tilde p$
\begin{equation}
\tilde p=-\tau.
\end{equation}
The metric $h^\mu_{\pht{\mu}\nu}$ on the time-like hypersurface $n=0$
created by the history of the membrane is given by equation \eqref{eq:shell_metric}.
The scalar $n$ measures the geodesic distance from this surface $n=0$.  
The divergence for the shell tensor $I_{\mu\nu}$ gives the two ``field
equations''
\begin{gather}
u^\mu u^\nu K_{\mu\nu}\big|^+_- +K^\lambda_\lambda\big|^+_-=-2\kappa\tau\\
u^\mu u^\nu K_{\mu\nu}\big|^+_-=\frac{1}{2}\kappa(\sigma-2\tau).
\end{gather}
They correspond in Newtonian theory to equations \eqref{eq:newtonian_tau}
and \eqref{eq:newtonian_F}.

\section{}
We sketch here another model that can be used to demonstrate Tolman’s 
paradox while taking the gravitational field fully into account. For 
this purpose we consider a solution of the Tolman-Oppenheimer-Volkoff 
equation for a spherically symmetric star consisting of radiation that 
is regular in the origin at $r=0$. The equation of state is then given 
by $p=\rho/3$. Bernd Schmidt and Alan Randall showed \cite{SR} that 
such solutions exist with the property that their pressure $p$ goes 
to zero when the radius $r$ goes to infinity. While such finite light 
stars do not exist we can consider a finite piece of radius $R$ and 
stabilize it with a membrane of suitable energy density and surface 
tension using the methods of the preceding paper.

Provided that $GM/R<4/9$ for mass $M$ and radius $R$ of the solution 
it has the same mass and radius as a Schwarzschild ball with those parameters. 
One can then imagine that such a Schwarzschild ball with constant 
energy density and vanishing pressure at its surface can be transformed 
quasi-statically into the previously constructed radiation ball including 
membrane.

\end{document}